\def\asec{\ifmmode ^{\prime\prime}\else$^{\prime\prime}$\fi}

\def\msunyr{\mbox{\,${\rm M_{\odot}\, yr^{-1}}$}}

\def\Mdot{\dot M}

\def\degs{\ifmmode ^{\circ}\else$^{\circ}$\fi}
\def\EE#1{\times 10^{#1}}
\def\kms{\mbox{\,km~s$^{-1}$}}
\def\lsim{\!\!\!\phantom{\le}\smash{\buildrel{}\over
 {\lower2.5dd\hbox{$\buildrel{\lower2dd\hbox{$\displaystyle<$}}\over
                                 \sim$}}}\,\,}
\def\gsim{\!\!\!\phantom{\ge}\smash{\buildrel{}\over
{\lower2.5dd\hbox{$\buildrel{\lower2dd\hbox{$\displaystyle>$}}\over
                               \sim$}}}\,\,}

\documentclass[12pt,preprint]{aastex}      %% PREPRINT:  a one-column, single-spaced document:

\usepackage[latin1]{inputenc}
\usepackage{lscape}

\slugcomment{Printed: \today}

\shorttitle{Disentangling the Nature of the Radio Emission in Wolf Rayet Stars}
\shortauthors{Montes et al.}

\begin{document}

\title{Disentangling the Nature of the Radio Emission in Wolf Rayet Stars}

\author{Gabriela~Montes\altaffilmark{1,2}, Miguel~A.~P\'erez-Torres\altaffilmark{1}, Antonio~Alberdi\altaffilmark{1}, and~Ricardo~F.~Gonz\'alez\altaffilmark{2} }
\altaffiltext{1}{Instituto de Astrof\'{\i}sica de Andaluc\'{\i}a, (IAA-CSIC), Camino
Bajo de Hu\'etor 50, E-18008 Granada, Spain; gmontes@iaa.es, torres@iaa.es,
antxon@iaa.es}

\altaffiltext{2}{Centro de Radioastronom\'{\i}a y Astrof\'{\i}sica UNAM,
Apartado Postal 3-72 (Xangari), 58089 Morelia, Michoac\'an, M\'exico;
g.montes@astrosmo.unam.mx, rf.gonzalez@astrosmo.unam.mx}

%\received{\today}

\begin{abstract}

We present quasi-simultaneous, multi-frequency VLA observations at 4.8, 8.4, and 22.5~GHz, of a sample of 13 Wolf Rayet (WR) stars, aimed at disentangling the nature of their radio emission and the possible detection of a non-thermal behavior in close binary systems. We detected 12 stars from our sample, for which we derived spectral information and estimated their mass loss rates. From our data, we identified four thermal sources (WR 89, 113,
138, and 141), and three sources with a composite spectrum (similar contribution of thermal and non-thermal emission; WR 8, 98, and 156). On the other hand, from the comparison with previous observations, we confirm the non-thermal spectrum of one (WR~105), and also found evidence of a composite spectrum for WR 79a, 98a, 104, and 133. Finally, we discuss the possible scenarios to explain the nature of the emission for the observed objects.

\end{abstract}


\section{INTRODUCTION}

Wolf Rayet (WR) stars are evolved, massive stars that are losing their mass rapidly through strong stellar winds (Conti 1976). In this scenario, hot, massive OB stars are considered to be the WR precursors that lose their external layers via stellar winds, leaving exposed their He-burning nuclei and H-rich surfaces. The stellar winds of  WR stars display high mass-loss rates [$\Mdot\approx (2$-$5)\EE{-5}$\msunyr] (e.g. Abbott et al. 1986, hereafter AB86; Crowther et al. 1995; Leitherer et al. 1995, 1997, hereafter LC95, LC97) and high terminal velocities [$v \approx (1000-2000)$\kms] (e.g. Crowther et al. 1995).
These stellar winds reveal themselves at several frequency ranges, e.g., tracing the terminal velocity in the P Cygni profiles in UV (Crowther et al. 1995), or showing an excess of emission at IR and radio frequencies (e.g. AB86, Contreras et al. 2004).

At radio frequencies, the excess of emission is associated with the contribution of the free-free thermal emission coming from the ionized and expanding envelope formed by the stellar wind. Early studies by Panagia \& Felli (1975; hereafter PF75), and Wright \& Barlow (1975; hereafter WB95) found that the thermal radio emission of an ionized, spherically symmetric, steady wind with an electron density profile, $n_e \propto r^{-2}$ ($r$ being the distance from the central star), follows a power-law with frequency, $S_\nu \propto \nu^\alpha$, with a positive spectral index $\alpha \approx 0.6$ (hereafter this stellar wind is referred to as ``standard'' wind). These authors also derived simple, useful relations between observable quantities, the flux density $S_\nu$ and the mass-loss rate $\dot{M}$ of the star. 
On the other hand, negative and flat ($\alpha\sim 0$) spectral indices are also observed in massive stars. These spectral indices are thought to be an indication of a composite spectrum, with a thermal plus a non-thermal contribution. The non-thermal emission is thought to be synchrotron emission produced by relativistic electrons, which have been accelerated either in strong shocks within the stellar wind, for single stars (White 1985), or in a wind-collision region (WCR) between the stellar components for binary systems (WR+0;  Usov 1992, Eichler \& Usov 1993). 
Therefore, radio observations can to determine both $\Mdot$ and, through simultaneous multi-frequency observations, the spectral index $\alpha$, which characterizes the emission process. 

In this way, observations can allow to identify the presence of both thermal, and non-thermal emission in WR stars (AB86, LC95, LC97, Chapman et. 1999, Contreras et al. 2004, Cappa et al. 2004).
For thermal sources, deviations from a standard wind, could result in the spectral index being either steeper ($>$0.6; Nugis et al. 1998) or slightly flatter than 0.6 (WB95). On the other hand, for non-thermal sources it is unclear whether the conditions to produce synchrotron emission are present in all of these stars. Recently, theoretical (Van Loo 2005) and observational (Dougherty \& Williams 2000, hereafter DW00) studies suggest that non-thermal sources require a binary companion. For the synchrotron emission to be detected, this must be produced within the optically-thin region of both stellar winds (Pittard et al. 2006), or there must be some low-opacity lines-of-sight in the wind through which we are observing the emission (DW00). For wide binary systems (P$>$1~yr), the non-thermal detection of the WCR has been supported by theoretical (Dougherty et al. 2003, Pittard et al. 2006) and observational studies (e.g. WR~147, Williams et al. 1997; WR~140, Dougherty et al. 2005). However, for close (P$<$1~yr) binary systems, it is unlikely that the emission from the WCR could be detected, since it is expected to be absorbed by the dense WR stellar wind (DW00).

In this paper, we present simultaneous, multi-frequency observations of a sample of 13 WR stars, using the VLA at 4.8, 8.4, and 22.5~GHz, aimed at disentangling the origin of their stellar wind radio emission through the analysis of their spectral index and time variability by comparison with previous observations. 
In particular, we focus on the possible detection of the WCR emission in close binary systems, and present new observations of two single stars previously classified as non-thermal sources.
In Section 2, we describe our observations. In Section 3, we show our results,  classification of emission nature, and mass-loss rate, $\Mdot$, determinations for each of the stars detected in our sample. 
We discuss our results in Section 4, describing common scenarios for the sources with similar spectrum behavior. Finally in Section 5 we summarize of our main results.

\section{OBSERVATIONS}

We performed radio observations of a sample of 13 WR stars, listed in Table \ref{tabla1}, with the Very Large Array (VLA) of the National Radio Astronomy Observatory (NRAO)\footnote{The NRAO is a facility of the National Science Foundation operated under cooperative agreement by  Associated Universities, Inc.}\ . 
Two sets of observations were performed in 2007 while the VLA was in D configuration: we observed five sources on April 21 at 4.8 and 8.4~GHz, and three additional sources on May 6 at 4.8, 8.4, and 22.5~GHz.
In March 2008, we observed seven sources with the VLA in C configuration: five of them were new sources  and the other two were also observed in 2007 (WR~79a and WR~105).

Data editing and calibration were carried out using the NRAO
Astronomical Image Processing System (AIPS) package, following
standard VLA procedures.  Absolute flux calibration was achieved by
observing 3C~48 and 3C~286.  The phase calibrators used for each
observation and their flux densities at each of the observing
frequencies, as well as the uncertainties from the calibration, are
given in Table \ref{tcal}.  For WR~105, self-calibration at 22.5~GHz
allowed us to improve the phase calibration of this source. However,
for the rest of the sources self-calibration was not possible because
of the weakness of the sources.  Expected off-source rms of $\sim50$
and $\sim40\mu$Jy/beam at 4.8 and 8.4~GHz respectively, and
$\sim60\mu$Jy/beam at 23~GHz, were successfully achieved for most of
the sources, with an on-source time of 15 minutes at 4.8 and 8.4~GHz,
and of 30 minutes at 22.5~GHz.  However, the observations of WR~79a,
89, 104, and 105, which were obtained with the VLA in D configuration,
suffered from sidelobe contamination, resulting in an increase of the
rms noise. We improved the rms of those images by excluding the short
baselines in the AIPS-IMAGR (UVRANGE=3,0).
We used the AIPS-IMFIT procedure in order to fit the position and flux
density of the detected sources. Since the sources are not expected to
be resolved, we fixed in IMFIT the standard parameters for unresolved
sources. We took a box size approximately twice the beam size of each
image. This procedure fits a 2D-Gaussian to the sources and estimates
the position of the maximum, the total flux density, and the errors of these
determinations. The error estimated by IMFIT considers both the
quality of the fit and the image rms.

We present in Table \ref{tfluxSI} the position, total flux density,
and spectral indices obtained for the 12 detected sources. The flux
density uncertainties were calculated considering both the IMFIT error
estimates and 2\% calibration uncertainties at 4.8 and 8.4~GHz, and
5\% at 22.5~GHz.  Note that the rms noise of the image is the dominant contribution to
the error budget.  The upper limits of the undetected sources were
taken as three times the rms noise.

We also used unreported archival VLA data for WR~105 and WR~133,
searching for a possible variability in their radio emission which
would indicate a likely non-thermal origin.  For WR~105 we used VLA
observations in B configuration taken on November 20, 1999 at 1.4,
4.8, and 8.4~GHz, and on November 27, 1999 at 4.8, 8.4 and 15~GHz.
For WR~133, we analyzed observations on May 31, 1993 at 1.4, 4.8, and
8.4~GHz, when the VLA was in BC configuration.  We should note that
the VLA configuration was not relevant for our purpose, since we were
interested only in the flux density of the stars, which are not
resolved at any VLA configuration.  We calibrated these archival
data in a similar way than our observations, reaching similar
values for the rms noise. For WR~105, self-calibration could be
performed at all the observed frequencies.

\section{RESULTS}

\subsection{Spectral Index Determinations}

We observed a total of 13 WR stars and detected 12 of them at least at
one frequency. In Table \ref{tfluxSI} we present the flux densities of
the detected sources at each observed frequency, as well as upper
limits for the undetected sources.  All the sources were observed
quasi-simultaneously at different frequencies on the same day, six of them observed in this
way for the first time (WR~8, 98, 113, 138, 141 and 156).  Owing to the
observed flux density variability for this kind of sources, we point
out the importance of the simultaneity in order to correctly determine
their spectral index.  Using this approach, we derived spectral
indices for the eight sources detected at all observing
frequencies.  We also derived lower limits for the spectral indices of
the remaining four sources that were detected at least at one
frequency, but with only an upper limit for the flux density at another.
When only two frequencies were available, the spectral index was
determined using the logarithmic ratio of the flux densities and the
observing frequencies. Otherwise, when three frequencies were
available, we used a linear regression fit weighting each data point
by its error.  For WR~133 we have not used the data at the lower
frequencies to determine its spectral index.  
The archival data at 1.4~GHz had a large rms noise, which
prevented us from fixing a reliable upper limit for its flux density.
On the other hand, our 4.8~GHz data did not allow us to determine an
upper limit to the flux density, since WR~133 is diluted by the
unresolved emission of a source to the northeast of the WR~133
position.

\subsection{Disentangling the Nature of the Emission}

The thermal stellar winds are expected to have spectral indices $\sim0.6$
(WB75, Nugis et al. 1998).  On the other hand, in non-thermal WR stars,
 the contribution of the thermal emission from the stellar winds combined with a non-thermal component from the WCR,
will result in a total spectrum characterized by spectral indices
$<0.6$ (e.g. Eichler \& Usov 1993).  This non-thermal emission is expected to be modulated
by the orbital motion of the system, resulting into variability of the
total spectrum (e.g. WR~140; Dougherty et al.  2005) with a
periodicity related with the orbital period.

In order to classify the nature of the WR radio emission, we have
determined i) the spectral index for each source and observing epoch
and ii) the presence of variability by comparing all the flux density
measurements from the different observing epochs.  We considered the
existence of flux density variability when the differences in the flux
densities between two or more epochs are higher than $3\sigma$. For
the spectral index, we use a conservative $3\,\sigma$ criterion as a
significant difference.  Optically-thin thermal emission (for which
$\alpha\sim -0.1$) is not expected for these massive stars at these
frequencies (DW00). In this way, we classified our sources as follows:

\begin{itemize}
\item Thermal (T; free-free thermal emission): sources with a spectral index $\gsim0.6$ for all the observing epochs.
\item Non-thermal (NT; dominant non-thermal emission): sources that showed a spectral index $\lsim-0.1$ for at least one observing epoch, as well as variability in their emission.
\item Composite (T/NT; thermal+non-thermal): sources that presented spectral indices higher than -0.1 but flatter than 0.6 (for at least one observing epoch), as well as variability in their emission.
\end{itemize}

Applying these criteria to the spectral indices and flux densities determined from these and previous observations, we found that at least four sources are T (WR~89, 113, 138, and 141), one is NT (WR~105), and three present a T/NT behavior (WR~8, WR~79a, and 156).

For the rest of the sources we have relaxed those criteria (WR~98, 98a, 104, and 133), since their strict application would mask important characteristics of these sources. For WR~98a and WR~104, we have also considered the modeling of the emission presented by Monnier et al. (2002), and classified them as T/NT sources.  For WR~98, the first determination of the spectral index is not conclusive about its spectrum; therefore, we also have taken into account the short time variability (15~days, see Table \ref{tcomp} ) at 4.8~GHz as a plausible indication of a possible non-thermal emission and a tentative T/NT nature. For WR~133, it is difficult to infer a variability in the spectrum from the two spectral index determinations, since those were determined from different frequency ranges (see Table \ref{tcomp} ). However, we should point out that the lower frequency range of WR~133 shows a tendency towards a non-thermal spectrum in the 1993 epoch.  Therefore, we classified WR~98 and WR~133 as tentative T/NT sources.

Summarizing, we have found four T (WR~89, 113, 138, and 141), one NT (WR~105), and seven T/NT sources (WR~8, 79a, 98, 98a, 104, 133, and 156).

\subsection{Mass-Loss Rates}

As we mentioned in Section 1, it is possible to estimate the free-free
radiation emitted from ionized extended envelopes.  PF75 and WB75
assumed a ``standard wind'' (steady and completely ionized, with
electron density profile $n_e\propto r^{-2}$, and expansion velocity
$v_\infty$), and showed that its flux density at radio frequencies
follows a power law with frequency as $S_{\nu}\propto \nu^{0.6}$.
Furthermore, those authors derived a general expression for the
mass-loss rate, $\dot M$, in terms of observable quantities, 
\begin{eqnarray}
\Bigg[\frac{\Mdot}{\rm{M_\odot \, yr^{-1}}}\Bigg]=5.34\times10^{-4}
\Bigg[\frac{S_\nu}{\rm{mJy}}\Bigg]^{3/4}
\Bigg[\frac{v_\infty}{\rm{km\,s^{-1}}}\Bigg]
\Bigg[\frac{d}{\rm{kpc}}\Bigg]^{3/2}
\Bigg[\frac{\nu}{\rm{Hz}}\Bigg]^{-1/2}
\Bigg[\frac{\mu^2}{Z \gamma g_\nu}\Bigg]^{1/2},
\label{mlr}
\end{eqnarray}
where $v_\infty$ is the terminal velocity of the stellar wind, $d$ is the distance to the star, $\nu$ is the observed frequency, and $g_\nu$ is the free-free Gaunt factor. The parameters $\mu$, $Z$ and $\gamma$ are the mean molecular weight, the average ionic charge, and the mean number of electrons per ion, respectively.

In Table \ref{tMLR}, we present $\Mdot$ determination using equation
(\ref{mlr}) for the 12 sources of our sample detected at 8.4~GHz.
Those determinations were done by assuming a standard stellar wind
with a filling factor, $f=1$; however, deviations from such
assumptions might lead into
an overestimation of $\dot M$. 

For the sources identified as NT and T/NT, the additional contribution to the thermal stellar wind emission would cause an overestimation for the $\dot M$ values. On the other hand, for the T sources, although their  spectral indices higher than 0.6 suggest an overestimation for $\dot M$, their impact is not expected to be significant at this frequency (Montes et al. in preparation). Therefore, these determinations can be considered reliable, at least for the T sources.

The values $\mu$, $Z$, and $\gamma$ were taken from CG04. For WR~138 and WR~141 we adopted the values of WR~133, since they share the same spectral type, WN5.
We compute $g_\nu$ from equation (3) within Leitherer \& Robert (1991), adopting $T_e$= 10$^4$~K (deviations from this temperature have minor effects on $g_\nu$).
The terminal velocities $v_{\infty}$ and distances $d$ of the stars were obtained from the van der Hucht (2001) WR catalog.
We estimated the uncertainties for each parameter in Table \ref{tMLR} using the same criteria as in LC97. 
For the parameter $d$ we assumed a 20\% error based on the cluster/association membership of the star, otherwise 40\% errors are assumed (0.09 and 0.25~dex, respectively). 
For the errors in $v_{\infty}$ we assume 10$\%$ (0.04~dex). The error obtained in $g_{\nu}$ is about 10$\%$. 
Finally, for $\mu$, $Z$, and $\gamma$ we assumed an error of $\pm$0.08~dex, as in CG04.
Therefore, following the equation (5) in LC97 for the $\Mdot$ uncertainties, we
obtained a typical logarithmic error of 0.21 for the stars considered in cluster/association and 0.41 for the rest (WR~98, 98a, 104, and 156).

We compared our results for $\Mdot$ with those previously reported by Cappa et al. (2004), we calculated  the average $\Mdot$ for each of the main spectral types in the sample, $\Mdot(\rm{WN})=(3.05\pm2.21)\times10^{-5}\,\rm{M_\odot \, yr^{-1}}$ and $\Mdot(\rm{WC})=(3.07\pm2.73)\times10^{-5}\,\rm{M_\odot \, yr^{-1}}$, for WN and WC respectively. We compared and found these values to coincide within the errors with those presented by Cappa et al. (2004; $\Mdot(\rm{WN})\sim4\times10^{-5}\,\rm{M_\odot \, yr^{-1}}$ and $\Mdot(\rm{WC8-9})\sim2\times10^{-5}\,\rm{M_\odot \, yr^{-1}}$  ). Nevertheless, this coincidence must be taken carefully due to the expected overestimation of $\Mdot$ mentioned above.

\section{DISCUSSION}

The results of our observations presented in Section 3 provide relevant information about the nature of the radio emission of the 12 detected WR stars. The detected flux densities and spectral indices displayed by the sources of our sample indicate the existence of thermal, non-thermal dominant, and composite spectrum sources.

For the sources identified as thermal (WR~89, 113, 138, and 141), we found spectral indices $\sim 1$.  
In a single star context, when the radio-continuum spectrum is consistent with thermal emission, the deviations of the expected 0.6 value for an isotropic wind (PF75 and WB75) may be caused either by the presence of condensations (clumps) that produce non-standard electron density profile ($n \propto r^{-s}$; with $s \neq$ 2), or by internal shocks which result from variations in the wind parameters at injection (see Gonz\'alez $\&$ Cant\'o 2008). 
Specifically, spectral indices higher than 0.6, may be the result of density profiles that fall off more rapidly than $n \propto r^{-2}$. 
On the other hand, for close binary systems, although the non-thermal emission from the WCR is expected to be absorbed by the WR stellar wind, the hot and dense shocked gas within the WCR might be able to contribute to the thermal radio spectrum (Stevens et al. 1995, Pittard et al. 2006, Pittard 2009).
 Moreover, in these kind of systems, it is possible that the WCR turns radiative, resulting in a dense and optically-thick thermal emitting structure with a positive spectral index ($\gsim 1$) (Pittard 2009, Montes et al. in preparation).
The nature of the WCR can be quantified through the ratio of the cooling time-scale for the shocked gas to the dynamical time-scale for it to flow out of the system  $\chi  \approx v^4_8 D_{12}/\dot{M}_{-7}$, where $v_8$ is the pre-shock velocity in units of  $10^8\rm{cm\,s^{-1}}$, $D_{12}$ is the stellar separation in units of  $10^{12}\,$cm, and $\dot{M}_{-7}$ is the stellar mass-loss rate in units of $10^{-7}\,\rm{M_\odot \,yr^{-1}}$ (Stevens et al. 1992).
 In this way, the WCR can be either radiative ($\chi<1$) or adiabatic ($\chi \gsim 1$). 
For the binary thermal sources in our sample, $D<1$~AU which implies $\chi < 0.5$, according to the stellar wind parameters given in Table \ref{tMLR}. 
This suggests a radiative WCR with a thermal contribution to the spectrum, which might be able to turn its spectral index into such positive values.
%This suggest a a radiative shock contributing to the thermal spectrum, turning  its spectral indices into such positive values.  
Therefore, for WR~113, 138 and 141, although their high spectral indices ($\sim1$) might result from a clumpy single stellar wind, it is also possible that such spectral indices are also indicating a binary star system.  
Moreover, for WR~141 there seems to be evidence for the presence of a WCR region from observations at other frequencies (X-rays, Oskinova 2005, and spectral analysis, Moffat et al. 1996).

On the other hand, the non-thermal indications of the single stars WR~79a (T/NT) and WR~105 (NT) have no clear explanation.
A non-thermal component in WR stars seem to be closely related to the binary nature of the star. DW00 analyzed the radio emission of a sample of 23 WR stars.
They found that most of the sources with a non-thermal signature were binary systems, which suggests that binarity is intrinsically related to the detection of non-thermal emission.
 In this scenario non-thermal emission is thought to be produced within a WCR between the two stellar components.
However, WR~79a and WR~105 do not show any evidence of binarity (Marchenko et al 1998).
Van~Loo et al. (2006) presented a study of the non-thermal emission produced by wind-embedded shocks resulting from instabilities within a single O-type stellar wind (also applicable to WR stars). They found that the radial decline of the shock strength (velocity jump and compression ratio) produces a rapid decrease of the synchrotron emission, becoming negligible at those radii where the stellar wind becomes optically-thin. This rules out the detection of a non-thermal component in a single star.
%Another alternative could be that thermal emission arising from this kind of shocks is strong enough to contribute to the total observed spectrum, as was proposed by Gonz\'alez \& Cant\'o (2008) for P~Cygni. However, it is unlikely that such contribution results in a spectrum with a negative $\alpha$, as that one already observed WR~105.
Moreover, as pointed out by DW00, proving that a star is single is very difficult, and we cannot rule out the presence of a companion in these stars.  
Since detectable non-thermal emission depends, in part, on the binary separation and the inclination angle of the system (e.g. WR~140 Dougherty et al. 2005), the detected flux density is expected to be modulated by the orbital motion, showing variability with a periodic behavior. Therefore, the observed variability in the spectra of WR 79a and WR 105 might be an indication of binary systems. Observations of WR 105 show variability in its 4.8 GHz flux density over a period of time of $\sim$ 14 yr. For WR 79a the emission at 8.4 GHz increases $\sim$ 50$\%$ between 2001 and 2005 observations (see Table 4). However, shorter variability time-scales in those sources can not be ruled out, as observations do not allow for a precise determination of their variability time scale and orbital period.

The tentative T/NT sources WR 98 and WR 133, are close binaries with periods of 47.8 and 112.4 days, respectively. Assuming a total mass for the system $M =$ 50 M$_{\odot}$, the semi-major axis of the orbits are $\sim$ 1 AU and 1.7 AU, respectively. Non-thermal emission arising from a WCR in these systems must be within the optically-thick region of the WR stellar wind ($\gsim30$~AU at 4.8~GHz; from equation (25) in PF75) and then not detectable. This seems difficult to reconcile with the tentative T/NT behavior displayed by WR~98 and WR~133. In fact, the short periods of these systems make it unlikely that the separation between the components ($D$) increases (even for a high eccentricity), at least to those radii required for non-thermal emission to escape the absorption ($\sim30$~AU).
Pittard et al. (2006) analyzed the thermal emission arising from non-radiative WCR (according to Stevens et al. 1992), and pointed out that thermal emission from the WCR may increase when  $D$ decreases (see Figure 11 in Pittard et al. 2006), contributing significantly to the total thermal flux, and resulting in a spectrum that may mimic a non-thermal contribution. 
In this scenario, an enhancement of 40$\%$ in the flux density (at 4.8 GHz) of the system WR 98 would suggest a decrease of 29$\%$ in the separation $D$ (since the thermal emission from the WCR is expected to scale as $D^{-1}$).
However, a WCR within this system would be roughly isothermal ($\chi\sim0.1$) with a positive thermal spectrum (Montes et al. in preparation). 
Therefore the tentative T/NT behavior for WR~98 is likely to be non-thermal emission from the WCR that is able to escape the absorption at certain orbital phases.
On the other hand, for WR~133 the WCR is likely to be adiabatic ($\chi\gsim 1$) with a flat thermal spectrum (see Pittard et al. 2006).
For WR~133 the presence of WCR is also supported by the strong emission at X-ray energies (Oskinova 2005) and optical spectroscopic analysis (Underhill \& Hill, 1994). However its influence at radio frequencies remains unclear, since the high errors of the spectral indices and the lack of observing epochs do not allow to infer any variability.
Detailed theoretical models are required to determine whether a spectrum like T/NT can be reproduced entirely by thermal emission or the non-thermal emission is able to be detectable for these systems.

WR~98a and WR~104 are well known colliding-wind binaries with an expanding dust spiral, called 'pinwheel nebulae' (Tuthill et a. 1999, Monnier et al. 1999, Tuthill et al. 2008). Monnier et al. (2002) modeled their radio emission taking into account a thermal plus an absorbed non-thermal component.
At 8.4~GHz, we detected WR~98a, but only marginally WR~104. However, we have not detected either of these sources at 4.8~GHz, probably as a consequence of the strong absorption of the non-thermal emission described by Monnier et al. (2002).
This absorption, and the lack of information about the flux density at higher frequencies, are likely to be the reason for the high value of the lower limits for their spectral indices.

Finally, although we classified WR~8 and WR~156 as T/NT, we can not completely rule out a single thermal wind origin of the observed emission.  The binarity of WR~8 have been questioned by Crowther et al. (1995). Moreover emission variability was only marginally determined from our criterion, and the uncertainty of its spectral index is high. The same occurs for the flux density and spectral index  determinations of WR~156. Further radio observations of both stars are needed to confirm its emission nature.

\section{SUMMARY}

We have presented simultaneous, multi-frequency observations of 13 WR stars at 4.8, 8.4, and 23~GHz. We have detected 12 of the observed sources at least at one frequency. The simultaneity of our observations have allowed us to obtain a reliable spectral index determination. In this way, we have determined spectral indices for eight, and lower limits for four of our 12 detected sources, being for six of them the first determinations. From the observed flux densities, spectral index determinations, and the comparison of our results with previous ones, we have disentangled the nature of the emission in these WR stars. 
We have also presented $\Mdot$ determinations, using our 8.4~GHz VLA flux density measurements. Those mass-loss rate estimates are consistent with values previously determined for WR stars.

We have classified WR~89, 113, 138, and 141 as thermal sources, for which $\alpha>0.6$.
We suggest different scenarios to reproduce the observed values for the spectral index. Such scenarios are mainly characterized by deviations from the standard wind, such as a different dependence with $r$ for the electron density profile ($s\neq2$), variability of both $\Mdot$ and $v_\infty$, and/or the influence of a binary companion.

WR~105, is likely a NT source owing to their significant flux density variability and the negative spectral index for at least one observing epoch. WR~8, 79a, 98, 98a, 104, 133, and 156 show indications for a composite spectrum (T/NT). 

The origin of the non-thermal signature for some of the sources of our sample is not completely clear, neither in the single stars (WR~79 and WR~105), nor in the close binary systems (WR~8, 98, and 133). 
The non-thermal signature in the WR 79a and WR 105 spectrum (thought to be single stars), could be explained by the presence of an undetected companion star. 
We also note the importance of carrying out a radio monitoring of the close binary systems WR 98 and WR 133. 
In fact, in case the flux density variability had a periodic behavior in those systems, and the period would happen to coincide with their orbital periods, this would unambiguously point to a binary origin of their NT signature.

\emph{Acknowledgements.} We thank an anonymous referee for useful comments. GM acknowledges financial support from CSIC predoctoral I3P and JAE fellowship. MAPT research is funded through a Ramon \& Cajal Fellowship of the Spain Ministry of Science (MEC) and the Spanish Research Council (CSIC). GM, MAPT and AA acknowledge support from the Spanish MEC through grant AYA 2006-14986-C02-01, and from the Consejer\'{\i}a de Innovaci\'on, Ciencia y Empresa of Junta de Andaluc\'{\i}a  through grants FQM-1747 and TIC-126.

\newpage

%--------------------------------------------------------------------------------------------------------------------------------------------
%                FIGURAS
%--------------------------------------------------------------------------------------------------------------------------------------------

%\begin{figure}
%\epsscale{1.0}
%\plotone{f1.eps}

%\caption{\footnotesize{Contour image for the WR stars detected at 4.8~GHz. Contour lines are -3, 3, $3\sqrt{3}$, 9, ... the rms noise. The rms noise is $\sim60~ \mu\rm{Jy}$, except for WR~89 and WR~105 for which is $\approx$3 times higher, owing to their low declinations and the influence of extended sources within the field of view. The synthesized beam is drawn at the bottom left of each image. 
%In each panel, the plus sign marks the optical positions of the star (we note that the size does not represent the uncertainty in the position). The cross in the WR~133 image marks the peak of the emission of the source detected at the northeast of WR~133 at 8.4~GHz.} }

%\label{Cfig}
%\end{figure}

\newpage

\begin{figure}
\epsscale{0.8}
\plotone{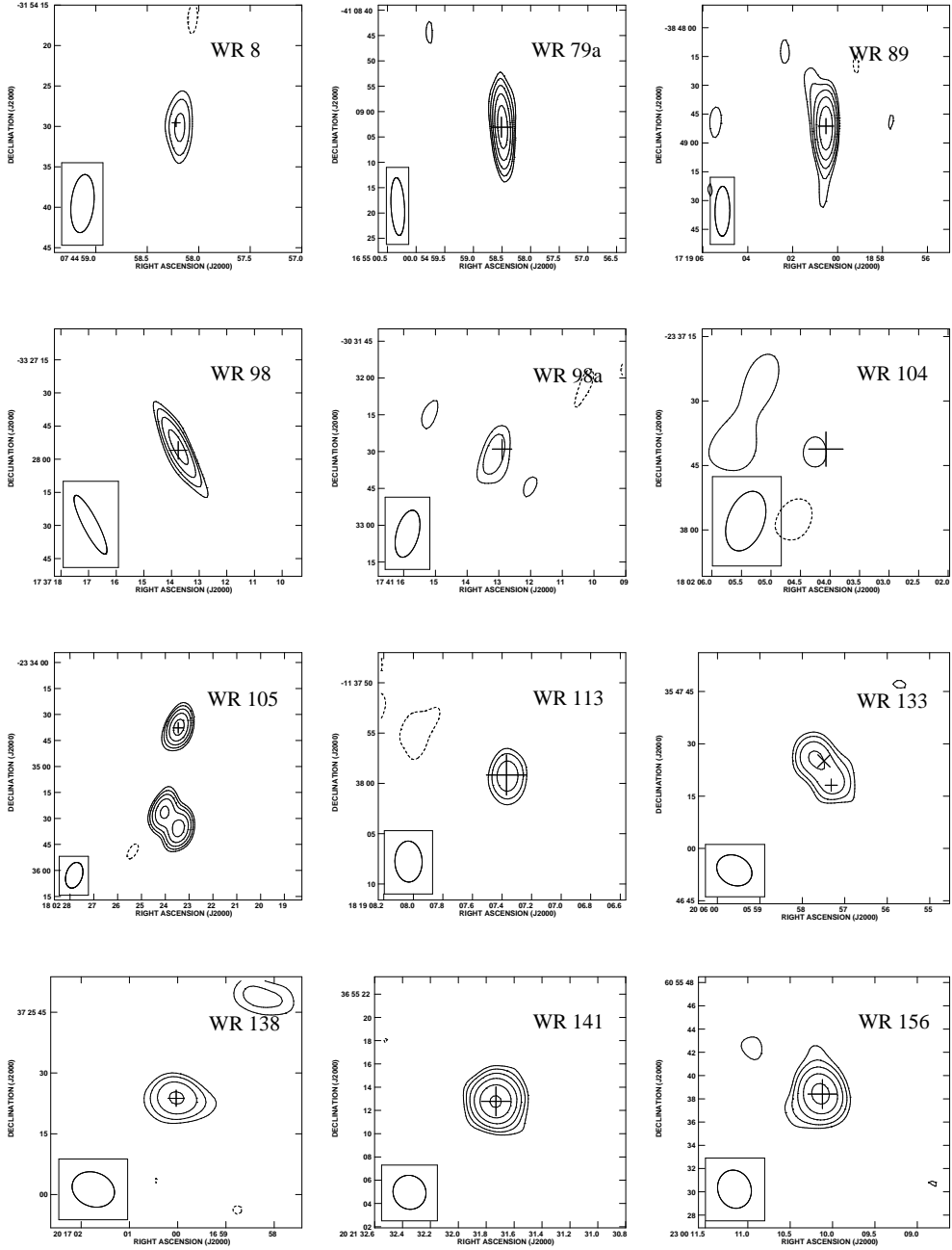}

\caption{\footnotesize{Contour image for the WR stars detected at 8.4~GHz of the 12 detected sources, including the tentative detection of WR~104. Contour lines are -3, 3, $3\sqrt{3}$, 9, ... the rms noise. The rms noise is $\sim50~ \mu\rm{Jy}$, except for WR~104 and WR~105 for which is $\approx$2 times higher, owing to their low declinations and the influence of extended sources within the field of view. The synthesized beam is drawn at the bottom left of each image.
In each panel, the plus sing marks the optical positions of the star (we note that the
size does not represent the uncertainty in the position). The cross-sign in the WR~133 image marks the peak of the emission at 4.8~GHz.}}

\label{Xfig}
\end{figure}

%\newpage

%\begin{figure}
%\epsscale{1.0}
%\plotone{f3.eps}

%\caption{\footnotesize{Contour image for the WR stars detected at 22.5~GHz. Contour lines are -3, 3, $3\sqrt{3}$, 9, ... the rms noise.  The rms noise is $\sim70~ \mu\rm{Jy}$ for all the sources. The synthesized beam is drawn at the bottom left of each image. In each panel, the plus sign marks the optical positions of the star (we note that the size does not represent the uncertainty in the position). The cross in WR~133 image marks the position of the emission peak of the source detected at the northeast of WR~133 at 8.4~GHz. The $3\sigma$ feature coinciding with this position suggest a marginal detection of this source.}}

%\label{Kfig}
%\end{figure}

\newpage

%--------------------------------------------------------------------------------------------------------------------------------------------
%                TABLAS
%--------------------------------------------------------------------------------------------------------------------------------------------

\begin{deluxetable}{clcc}
\tabletypesize{\scriptsize}
%\rotate
\tablewidth{0pt}
%\tablenum{}
\tablecaption{Wolf Rayet Sample and their Properties\label{tabla1}}
%\tablehead{}
%\tablecolumns{}
\startdata
\hline \hline

WR       &    Spectral Type   & Binary Status  &   P        \\
         &                    &                & (days)     \\
\hline
8        &  WN7/WCE+?         & SB1            & 38.4,115   \\
12       &  WN8h+?            & SB1,no d.e.l.  & 23.92      \\
79a      &  WN9ha             & VB             & ...        \\
89       &  WN9h+OB           & a.d.l.,VB      & ...        \\
98       &  WN8/WC7           & SB1            & 48.7       \\
98a      &  WC8-9vd+?         & CWB            & 565        \\
104      &  WC9d+B0.5V        & SB2,VB         & 243        \\
105      &  WN9h              & ...            & ...        \\
113      &  WC8d+O8-9IV       & SB2            & 29.7       \\
133      &  WN5+O9I           & SB2,VB         & 112.4      \\
138      &  WN5+B?            & SB2,VB         & 11.6,1538  \\
141      &  WN5+O5V-III       & SB2            & 21.6       \\
156      &  WN8h+OB?          & d.e.l.         & 6.5,10     \\
\enddata
\\

\tablecomments{The values displayed in the columns were taken from the van der Hucht (2001) WR catalog.}

\end{deluxetable}

\newpage

\begin{deluxetable}{cccccccccccccc}
\tabletypesize{\tiny}
%\rotate
\tablewidth{0pt}
%\tablenum{}
\tablecaption{Calibrators \label{tcal}}
%\tablehead{}
%\tablecolumns{}
\startdata

\hline \hline
Source & Observation     & Phase      & \multicolumn{5}{c}{Bootstrapped flux (Jy) \tablenotemark{a}}  \\
\cline{4-8} 
WR No. & Date & Calibrator & 1.4~GHz &  4.8~GHz    &     8.4~GHz     & 15~GHz & 23~GHz     \\
\hline
79a    & 07-Apr-21 & 1626-298 &$\cdots$  & 2.355$\pm$0.005 & 2.208$\pm$0.004 &$\cdots$ & $\cdots$        \\
89     & 07-Apr-21 & 1626-298 &$\cdots$  & 2.355$\pm$0.005 & 2.208$\pm$0.004 &$\cdots$ & $\cdots$        \\
98     & 07-Apr-21 & 1751-253 &$\cdots$  & 0.469$\pm$0.001 & 0.265$\pm$0.004 &$\cdots$ & $\cdots$        \\
98a    & 07-Apr-21 & 1751-253 &$\cdots$  & 0.469$\pm$0.001 & 0.265$\pm$0.004 &$\cdots$ & $\cdots$        \\
104    & 07-Apr-21 & 1820-254 &$\cdots$  & 0.630$\pm$0.005 & 0.695$\pm$0.002 &$\cdots$ & $\cdots$        \\
105    & 07-Apr-21 & 1820-254 &$\cdots$  & 0.630$\pm$0.005 & 0.695$\pm$0.002 &$\cdots$ & $\cdots$        \\
98     & 07-May-06 & 1744-312 &$\cdots$  & 0.570$\pm$0.005 & 0.637$\pm$0.003 &$\cdots$ & 0.590$\pm$0.006 \\
133    & 07-May-06 & 2025+337 &$\cdots$  & 1.701$\pm$0.003 & 2.275$\pm$0.006 &$\cdots$ & 2.657$\pm$0.016 \\
138    & 07-May-06 & 2025+337 &$\cdots$  & 1.701$\pm$0.003 & 2.275$\pm$0.006 &$\cdots$ & 2.657$\pm$0.016 \\
\hline
8      & 08-Mar-07 & 0735-175 &$\cdots$  & 1.527$\pm$0.033 & 0.954$\pm$0.009 &$\cdots$  & 0.429$\pm$0.006 \\
12     & 08-Mar-07 & 0735-175 &$\cdots$  & 1.527$\pm$0.033 &  $\cdots$       &$\cdots$  &  $\cdots$	    \\
79a    & 08-Mar-05 & 1626-298 &$\cdots$  & 2.759$\pm$0.030 & 2.696$\pm$0.008 &$\cdots$  & 1.952$\pm$0.044 \\
113    & 08-Mar-05 & 1832-105 &$\cdots$  & 1.315$\pm$0.003 & 1.437$\pm$0.003 &$\cdots$  & 0.925$\pm$0.016 \\
105    & 08-Mar-05 & 1820-254 &$\cdots$  & 0.932$\pm$0.019 &  $\cdots$       &$\cdots$  & 0.877$\pm$0.016 \\
141    & 08-Mar-08 & 2015+371 &$\cdots$  & 1.475$\pm$0.010 & 2.056$\pm$0.010 &$\cdots$  & 3.298$\pm$0.073 \\ 
156    & 08-Mar-08 & 2148+611 &$\cdots$  & 1.360$\pm$0.003 & 1.068$\pm$0.004 &$\cdots$  & 0.714$\pm$0.016 \\
\hline
105    & 99-Nov-20 & 1820-254 & 1.163$\pm$0.009 &  0.948$\pm$0.003 & 0.864$\pm$0.004 & $\cdots$ & $\cdots$  \\
105    & 99-Nov-27 & 1820-254 & $\cdots$        &  0.953$\pm$0.002 & $\cdots$        & 0.778$\pm$0.0134 & $\cdots$\\
\hline 
133    & 93-May-31 & 2005+403 & 2.60$\pm$0.03   & 3.14$\pm$0.05 & 3.12$\pm$0.08                & $\cdots$ & $\cdots$  \\ 
\enddata
\tablecaption{Calibrators \label{tcal}}
\tablecomments{Lines separates the observing years.}

\scriptsize
\tablenotetext{a}{The bootstrapped fluxes and their uncertainties are those obtained from GETJY during the calibration.}

\end{deluxetable}

\clearpage

\begin{deluxetable}{ccccccccccc}
\tabletypesize{\scriptsize}
%\rotate
\tablewidth{0pt}
%\tablenum{}
\tablecaption{Radio Flux Densities and Spectral Indices \label{tfluxSI}}
%\tablehead{}
%\tablecolumns{}
\startdata
\hline \hline
Source  & Observation &  \multicolumn{3}{c}{Positions at 8.4~GHz (J.2000.0)} & & \multicolumn{3}{c}{Fluxes (mJy)} &   & \\ \cline{3-5} \cline{7-9}
WR No.  &  Date       &R.A.    &  Dec.       &Error& &  4.8~GHz      & 8.4~GHz        & 23~GHz & $\alpha$\\

\hline

79a & 07-Apr-21 & 16 54 58.51 & -41 09 04.3 & 0.4 & & 0.86$\pm$0.07 & 1.67$\pm$0.06 &...            & 1.16$\pm$0.16\\

89  & 07-Apr-21 & 17 19 00.53 & -38 48 52.0 & 0.2 & & 1.82$\pm$0.24 & 3.01$\pm$0.08 &...            & 0.88$\pm$0.24  \\

98  & 07-May-06 & 17 37 13.82 & -33 27 55.4 & 0.3 & & 0.94$\pm$0.06 & 1.09$\pm$0.05 &...            & 0.26$\pm$0.14 \\

98a & 07-May-06 & 17 41 13.12 & -30 32 32.0 & 1.0 & & $<$0.39       & 0.59$\pm$0.06 &...            &$>$0.73         \\

104\tablenotemark{a}& 07-May-06 & 18 02 04.31 & -23 37 41.4 & 0.8 & & $<$0.21       & 0.42$\pm$0.09 & ...           &$>$1.22         \\                   

105 & 07-May-06 & 18 02 23.44 & -23 34 37.5 & 0.1 & & 2.51$\pm$0.11 & 4.38$\pm$0.12 &...            & 0.98$\pm$0.09  \\

98  & 07-May-06 & 17 37 13.75 & -33 27 54.9 & 0.4 & & 0.58$\pm$0.06 & 1.18$\pm$0.05 & 1.94$\pm$0.15 & 0.64$\pm$0.07    \\

133 & 07-May-06 & 20 05 57.25 & +35 47 18.2 & 0.4 & & $<$0.41       & 0.31$\pm$0.03 & 0.57$\pm$0.07 & 0.60$\pm$0.15    \\

138 & 07-May-06 & 20 17 00.01 & +37 25 23.7 & 0.2 & & $<$0.12       & 0.52$\pm$0.03 & 1.15$\pm$0.08 & $>$0.82       \\
\hline
8   & 08-Mar-07 & 07 44 58.19 & -31 54 30.0 & 0.3 & & 0.22$\pm$0.04 & 0.28$\pm$0.03 & 0.30$\pm$0.05 & 0.17$\pm$0.15       \\

12\tablenotemark{a} & 08-Mar-07 & 08 44 47.18 & -45 58 55.9 & 7.8 & & $<$0.22       & ...           & ...      &...               \\ 

79a & 08-Mar-05 & 16 54 58.49 & -41 09 03.0 & 0.1 & & 0.99$\pm$0.04 & 1.56$\pm$0.05 & 2.31$\pm$0.14 & 0.54$\pm$0.05 \\

105\tablenotemark{b} & 08-Mar-05 & 18 02 23.46 & -23 34 37.4 & 0.1 & & 4.22$\pm$0.13 & ...           & 7.63$\pm$0.39 & 0.38$\pm$0.04    \\ 

113 & 08-Mar-05 & 18 19 07.36 & -11 37 59.3 & 0.1 & & 0.22$\pm$0.03 & 0.47$\pm$0.04 & 1.27$\pm$0.07 & 1.06$\pm$0.09  \\

141 & 08-Mar-08 & 20 21 31.73 & +36 55 12.8 & 0.1 & & 0.59$\pm$0.04 & 1.28$\pm$0.04 & 2.86$\pm$0.15 & 0.91$\pm$0.05  \\

156 & 08-Mar-08 & 23 00 10.15 & +60 55 38.4 & 0.1 & & 0.77$\pm$0.04 & 0.99$\pm$0.04 & 1.59$\pm$0.09 & 0.46$\pm$0.05  \\
\enddata

\tablecomments{Upper limits of the undetected sources were fixed as three times the rms noise of the map. For WR~133 at 4.8~GHz the upper limit was fixed from the peak of the unresolved emission.  The $\alpha$ errors were estimated from, $\Delta\alpha^2=[(\partial \alpha/ \partial S_1) \Delta S_1]^2 + [(\partial \alpha/ \partial S_2) \Delta S_2]^2$, for two frequency determinations of $\alpha$, and from the error of the weighed linear regression fit used for the three frequency determinations. See Section 2 for details about the error of the flux densities determinations. Lines separate the observing epochs.}
\tablenotetext{a}{Probable detection. Coordinates refer to the local maximum position at 4.8~GHz for WR~12 and at 8.4~GHz for WR~104, both close to the optical positions.}\tablenotetext{b}{Position at 23~GHz.}
\end{deluxetable}

\begin{deluxetable}{ccccccccccc}
\tabletypesize{\scriptsize}
%\rotate
\tablewidth{0pt}
%\tablenum{}
\tablecaption{Radio Flux Densities and Spectral Index Comparison with Previous Observations\label{tcomp}}
%\tablehead{}
%\tablecolumns{}
\startdata
\hline \hline

WR   &  $S_{1.4\,\mathrm{GHz}}$  & $S_{2.4\,\mathrm{GHz}}$  & $S_{4.8\,\mathrm{GHz}}$    &   $S_{8.4\,\mathrm{GHz}}$ & $S_{15\,\mathrm{GHz}}$    & $S_{23\,\mathrm{GHz}}$ & $\alpha$ & Obs Date & Ref. & Spec. Type\\
     &  (mJy)         & (mJy)         &     (mJy)      &     (mJy)     &     (mJy)     &  (mJy)     &  &   &   \\

\hline\\

8    & ...    & ...         & ...           & 0.36$\pm$0.03 & ...           & ...           &     ...       & 01Nov12& CG04 &\\   
     & ...    & ...         & 0.22$\pm$0.04 & 0.25$\pm$0.03 & ...           & 0.30$\pm$0.05 & 0.17$\pm$0.15 & 08Mar05 & TS  & (T/NT)\tablenotemark{a}\\
\hline\\
12   & ...    & ...         & ...           & 0.51$\pm$0.06 & ...           & ...           &       ...     & 01Nov12& CG04 & \\ 
     & ...    & ...         & $<$0.22       & ...           & ...           & ...           &       ...     & 08Mar05& TS   & \\ 

\hline\\

79a  &   ...  & ...         & 1.1$\pm$0.1   & ...           &  2.4$\pm$0.1  & ...           & 0.9$\pm$0.1   & 84Apr03 & BA89 &  \\
     &  ...   & 1.0$\pm$0.1 & 0.8$\pm$0.1   & 0.9$\pm$0.1   & ...           & ...           &-0.1$\pm$0.1   &  Nov00  & SC03 &  \\
     &  ...   & ...         &  ...          & 0.7$\pm$0.1   & ...           & ...           &         ...   & 01Sep15 & CG04 &  \\
     &  ...   & ...         & 0.86$\pm$0.07 & 1.67$\pm$0.06 & ...           & ...           & 1.16$\pm$0.16 & 07Apr21 & TS   &  \\
     &  ...   & ...         & 0.99$\pm$0.04 & 1.56$\pm$0.05 & ...           & 2.31$\pm$0.14 & 0.54$\pm$0.05 & 08Mar05 & TS   & (T/NT)\tablenotemark{a}\\
\hline\\
89   &  ...   & ...         & 0.6$\pm$0.1   & ...           & ...           & ... &      ...      & 82Aug20 & AB86 & \\
     &  ...   & ...         & 1.94$\pm$0.19 & 2.99$\pm$0.10 & ...           & ... & 0.76$\pm$0.18 & 94Sep08 & LC95 & \\
     &$<$1.20 & $<$0.90     & ...           & ...           & ...           & ... &     ...       & 97Feb23 & CL99 & \\
     &  ...   & ...         &  ...          & 2.0$\pm$0.1   & ...           & ... &     ...       & 01Nov12 & CG04 & \\
     &  ...   & ...         & 1.82$\pm$0.52 & 3.01$\pm$0.09 & ...           & ... & 0.88$\pm$0.24 & 07Apr21 & TS   & (T)\tablenotemark{a}\\
\hline\\
98   & ...    & ...         & 0.9$\pm$0.07  &   ...         & ...           & ...  & ...           & 85Aug17 & AB86 &  \\
     & ...    & ...         & 0.94$\pm$0.06 & 1.09$\pm$0.05 & ...           & ...  & 0.26$\pm$0.14 & 07Apr21 & TS   &  \\
     & ...    & ...         & 0.58$\pm$0.06 & 1.18$\pm$0.05 & ...           &1.94$\pm$0.15 & 0.64$\pm$0.07   & 07May06& TS & (T/NT?)\tablenotemark{c}\\
\hline\\
98a  &$<$0.36 & ...         &0.37$\pm$0.07&0.60$\pm$0.05& 0.64$\pm$0.11 & 0.57$\pm$0.10 & 0.30$\pm$0.22 & 00Feb24& MT02    &  \\
     & ...    & ...         & ...         &0.47$\pm$0.05& ...           & ...           &          ...     & 01Oct08& CG04 &  \\
     & ...    & ...         &$<$0.39      &0.59$\pm$0.06& ...           & ...           &     $>$0.73 & 07Apr21& TS        &  (T/NT)\tablenotemark{b}\\ 
\hline\\ 
104  & ...    & ...         &$<$0.4       & ...         & ...           & ...           & ...             & 84Apr04  & AB86 &  \\
     &  ...   & ...         &$<$2.01      &$<$0.39      & ...           & ...           & ...             & 94Sep07  & LC97 &  \\
     &$<$1.59 & $<$0.99     & ...         & ...         & ...           & ...           & ...             & 97Feb23  & CL99 &  \\
     &$<$0.30 & ...         &  ...        &0.87$\pm$0.06& 1.02$\pm$0.12 & 0.94$\pm$0.10 & 0.10$\pm$ 0.23  & 00Feb25  & MT02 &  \\
     &  ...   & ...         &  ...        &0.54$\pm$0.06& ...           & ...           &            ...  & 01Nov12  & CG04 &  \\
     &  ...   & ...         &$<$0.21      &0.42$\pm$0.09&               &               &     $>$1.22     & 07Apr21  & TS   & (T/NT)\tablenotemark{b} \\
\hline\\
105  &  ...   & ...         & 3.6$\pm$0.2 & ...           & ...           & ...           & ...           & 84Apr04  & AB86 & \\
     &  ...   & ...         &4.39$\pm$0.15& 3.75$\pm$0.15 & ...           & ...           &-0.28$\pm$0.01 & 94Sep07  & LC97 & \\
     &$<$1.17 & $<$0.69     & ...         & ...           & ...           & ...           &          ...  & 97Feb23  & CL99 & \\
     &1.43$\pm$0.20& ...    &2.92$\pm$0.08& 4.77$\pm$0.10 & ...           & ...           & 0.81$\pm$0.04 & 99Nov20  & TS   & \\

     & ...    & ...         &2.73$\pm$0.07& ...           & 7.02$\pm$0.37 & ...           & 0.83$\pm$0.05 & 99Nov27  & TS   & \\

     &  ...   & ...         &  ...        &  5.4$\pm$0.1  & ...           & ...           &         ...   & 01Nov12  & CG04 &\\
     &  ...   & ...         &2.51$\pm$0.11& 4.38$\pm$0.12 & ...           & ...           & 0.98$\pm$0.09 & 07Apr21  & TS   &\\
     &  ...   & ...         &4.22$\pm$0.13& ...           & ...           & 7.63$\pm$0.39 & 0.38$\pm$0.04 & 08Mar05  & TS   & (NT) \tablenotemark{a}\\
\hline\\
113  &  ...   & ...         & $\leq$0.4     &...            & ...           & ...           & ...           & 80Jul26 & BA82 &\\
     &  ...   & ...         & $<$0.80       & $<$0.80       & ...           & ...           & ...           & 94Sep07 & LC97 &\\
     &$<$2.25 &$<$0.90      &  ...          & ...           & ...           & ...           & ...           & 97Feb23 & CL99 &\\
     &  ...   & ...         & ...           & 0.75$\pm$0.04 & ...           & ...           & ...           & 01Nov12 & CG04 &\\
     &  ...   & ...         & 0.22$\pm$0.03 & 0.47$\pm$0.04 & ...           & 1.27$\pm$0.07 & .  1.06$\pm$0.09 & 07May06   & TS & (T)\tablenotemark{a}\\  
\hline\\
133  &  ...    & ...         &$<$0.3       &   ...       & ...           & ...          &  ...          & 85Aug17 & AB86 &\\
     & $<$0.81 & ...         &0.38$\pm$0.08&0.27$\pm$0.03& ...           & ...          &-0.65$\pm$0.42 & 93May31 & TS   &\\
     &  ...    & ...         & ...         &0.36$\pm$0.03& ...           & ...          &  ...          & 01Nov12 & CG04 &\\
     &  ...    & ...         & $<$0.41     &0.31$\pm$0.03& ...           &0.57$\pm$0.07 & 0.60$\pm$0.15 & 07May06 & TS   & (T/NT?)\tablenotemark{d}\\
\hline\\
138  & ...    & ...         & 0.6$\pm$0.1   & ...         & ...          & ...          &     ...       & 80Jul27& BA82 &\\
     &  ...   & ...         & $<$0.12       &0.52$\pm$0.03& ...          &1.15$\pm$0.08 & $>$0.82       & 07May06& TS   & (T)\tablenotemark{a}\\  
\hline\\
141  & ...    & ...         & 0.6$\pm$0.1   & ...           & ...          & ...          &         ...   & 85Aug05 & AB86 &\\
     & ...    & ...         & 0.59$\pm$0.04 & 1.28$\pm$0.04 & ...          &2.86$\pm$0.15 & 0.91$\pm$0.05 & 08Mar05 & TS   & (T)\tablenotemark{a}\\
\hline\\
156  &  ...   & ...         & ...           & 1.06$\pm$0.03 & ...          & ...           & ...           & 01Nov12 & CG04& \\
     & ...    & ...         & 0.77$\pm$0.04 & 0.99$\pm$0.04 & ...          & 1.59$\pm$0.09 & 0.46$\pm$0.05 & 08Mar05 & TS  & (T/NT)\tablenotemark{a}\\

\enddata

\tablecomments{The whole frequency range observed was used for the spectral index and lower limits determinations, except for WR~133, for which the upper limits of the flux densities at the lower frequencies were not taken into account. \\
For all the sources spectral classification was done over the whole group of observations. (T) Thermal. (NT) Non-thermal. (T/NT) Composite, thermal+non-thermal. (T/NT?) tentative composite. \\Spectral classification criteria used: }
\tablenotetext{a}{According to the criteria described in section 3.2.}
\tablenotetext{b}{Study presented in MT02}
\tablenotetext{c}{According to the criteria described in section 3.2 and the short time variability  $\sim$15~days in the density flux.}
\tablenotetext{d}{According to the criteria described in section 3.2 and the flat tendency of the spectral index determination at the lower frequency range.}
\tablerefs{TS:This Study; BA82: Bieging et al. (1982); BA89: Bieging et al. (1989); SC03: Setia Gunawan et al. (2003); MT02: Monnier et al. (2002).}

\end{deluxetable}

\begin{deluxetable}{clccccccr}
\tabletypesize{\scriptsize}
%\rotate
\tablewidth{0pt}
%\tablenum{}
\tablecaption{Mass-Loss Rate Determinations\label{tMLR}}
%\tablehead{}
%\tablecolumns{}
\startdata
\hline \hline

WR       & Spectral Type & $v_\infty$            &  $d$   & $\mu$ & $Z$  & $\gamma$  & $S_{8.4~\rm{GHz}}$& $\dot{M}_{8.4~\rm{GHz}}$    \\
         & & ($\rm{km\,s^{-1}}$)   &  (kpc) &       &      &           &  (mJy)            & $10^{-5}\,\rm{M_\odot\,yr^{-1}}$      \\
	   				      		   	                
\hline\\   				      		   	                
	   				      		   	                
8        & WN7/WCE+?     & 1590                  & 3.47   & 1.7   & 1.0  & 1.0       &  0.28  &   $<$1.79  \\
79a      & WN9ha         & 935                   & 1.99   & 2.6   & 1.0  & 1.0       &  1.56  &   $<$2.67       \\
89       & WN9h+OB       & 1600                  & 2.88   & 1.5   & 1.0  & 1.0       &  3.01  &   7.14      \\
98       & WN8/WC7       & 1200                  & 1.9    & 3.7   & 1.0  & 1.0       &  1.09  &   $<$3.30     \\
98a      & WC8-9vd+?     & 2000                  & 1.9    & 4.7   & 1.1  & 1.1       &  0.59  &   $<$4.03      \\
104      & WC9d+B0.5V    & 1220                  & 2.3    & 4.7   & 1.1  & 1.1       &  0.42  &   $<$2.54      \\
105      & WN9h          & 700                   & 1.58   & 2.6   & 1.0  & 1.0       &  4.38  &   $<$2.91      \\
113      & WC8d+O8-9IV   & 1700                  & 1.79   & 4.7   & 1.1  & 1.1       &  0.47  &   2.64      \\
133      & WN5+O9I       & 1800                  & 2.14   & 4.0   & 1.1  & 1.1       &  0.27  &   $<$2.06       \\
138      & WN5+B?        & 1400                  & 1.26   & 4.0   & 1.1  & 1.1       &  0.52  &   1.18       \\
141      & WN5+O5V-III   & 1550                  & 1.26   & 4.0   & 1.1  & 1.1       &  1.28  &   2.57       \\
156      & WN8h+OB?      & 660                   & 3.56   & 3.3   & 1.0  & 1.1       &  0.99  &   $<$3.87       \\
\enddata
\tablecomments{The logarithmic error in the $\dot{M}$ estimates are 0.21 for the stars considered in cluster/association, and 0.41 for the rest (WR~98, 98a, 104 and 156).}

\end{deluxetable}


\begin{references}

\reference{} Abbott, D.~C., Torres, 
A.~V., Bieging, J.~H., \& Churchwell, E.\ 1986, \apj, 303, 239 

\reference{} Annuk, K.\ 1990, Acta 
Astronomica, 40, 267 

\reference{} Bohannan, B., \& Crowther, P.~A.\ 1999, \apj, 511, 374 

\reference{} Bieging, J.~H., Abbott, 
D.~C., \& Churchwell, E.~B.\ 1982, \apj, 263, 207 

\reference{} Bieging, J.~H., Abbott, 
D.~C., \& Churchwell, E.~B.\ 1989, \apj, 340, 518 


\reference{} Cappa, C., Goss, W.~M., 
\& van der Hucht, K.~A.\ 2004, \aj, 127, 2885 

\reference{} Chapman, J.~M., 
Leitherer, C., Koribalski, B., Bouter, R., 
\& Storey, M.\ 1999, \apj, 518, 890 

\reference{} Cranmer, S.~R., \& Owocki, S.~P.\ 1994, Bulletin of the American Astronomical Society, 26, 1446 



\reference{} Conti, P.~S.\ 1976, Memoires of 
the Societe Royale des Sciences de Liege, 9, 193 


\reference{} Contreras, M.~E., 
Montes, G., 
\& Wilkin, F.~P.\ 2004, Revista Mexicana de Astronomia y Astrofisica, 40, 53 


\reference{} Crowther, P.~A., Smith, L.~J., \& Willis, A.~J.\ 1995, \aap, 304, 269 

\reference{} Dougherty, S.~M., \& Williams, P.~M.\ 2000, \mnras, 319, 1005 


\reference{} Dougherty, S.~M., Pittard, J.~M., Kasian, L., Coker, R.~F., Williams, P.~M., \& Lloyd, H.~M.\ 2003, \aap, 409, 217 

\reference{} Dougherty, S.~M., Beasley, A.~J., Claussen, M.~J., Zauderer, B.~A., \& Bolingbroke, N.~J.\ 2005, \apj, 623, 447 

\reference{} Eichler, D., \& Usov, V.\ 1993, \apj, 402, 271 


\reference{} Gamen, R.~C., \& Niemela, V.~S.\ 2002, New Astronomy, 7, 511 


\reference{} Gonz{\'a}lez, R.~F., \& Cant{\'o}, J.\ 2008, \aap, 477, 373 


\reference{} Hamann, W.-R., Gr{\"a}fener, G., \& Liermann, A.\ 2006, \aap, 457, 1015 



\reference{} Leitherer, C., \& Robert, C.\ 1991, \apj, 377, 629 

\reference{} Leitherer, C., Chapman, J.~M., \& Koribalski, B.\ 1995, \apj, 450, 289 




\reference{} Leitherer, C., 
Chapman, J.~M., \& Koribalski, B.\ 1997, \apj, 481, 898 

\reference{} L{\'e}pine, S., et 
al.\ 2000, \aj, 120, 3201 


\reference{} Mason, B.~D., Gies, 
D.~R., Hartkopf, W.~I., Bagnuolo, W.~G., Jr., ten Brummelaar, T., 
\& McAlister, H.~A.\ 1998, \aj, 115, 821 


\reference{} Marchenko, S.~V., 
Moffat, A.~F.~J., \& Eenens, P.~R.~J.\ 1998, \pasp, 110, 1416 




\reference{} Moffat, A.~F.~J., 
Drissen, L., Lamontagne, R., \& Robert, C.\ 1988, \apj, 334, 1038 


\reference{} Monnier, J.~D., 
Tuthill, P.~G., \& Danchi, W.~C.\ 1999, \apjl, 525, L97 

\reference{} Monnier, J.~D., 
Greenhill, L.~J., Tuthill, P.~G., \& Danchi, W.~C.\ 2002, \apj, 566, 399 

\reference{} Montes G., Gonz\'alez R.F., P\'erez-Torres M.A. \& Alberdi A. (in preparation)

\reference{} Niemela, V.~S.\ 1991, 
Wolf-Rayet Stars and Interrelations with Other Massive Stars in Galaxies, 
143, 201 

\reference{} Niemela, V.~S., Gamen, 
R., Morrell, N.~I., 
\& Gim{\'e}nez Ben{\'{\i}}tez, S.\ 1999, Wolf-Rayet Phenomena in Massive Stars and Starburst Galaxies, 193, 26 


\reference{} Nugis, T., Crowther, P.~A., \& Willis, A.~J.\ 1998, \aap, 333, 956 

\reference{} Oskinova, L.~M.\ 2005, 
\mnras, 361, 679 

\reference{} Panagia, N., \& Felli, M.\ 1975, \aap, 39, 1 


\reference{} Prinja, R.~K., Stahl, O., Kaufer, A., Colley, S.~R., Crowther, P.~A., \& Wolf, B.\ 2001, \aap, 367, 891 

\reference{} Pittard, J.~M., \& Dougherty, S.~M.\ 2006, \mnras, 372, 801 

\reference{} Pittard, J.~M., Dougherty, S.~M., Coker, R.~F., O'Connor, E., \& Bolingbroke, N.~J.\ 2006, \aap, 446, 1001 

\reference{} Pittard, J.~M.\ 2009, arXiv:0908.1003 




\reference{} Setia Gunawan, 
D.~Y.~A., Chapman, J.~M., Stevens, I.~R., Rauw, G., 
\& Leitherer, C.\ 2003, A Massive Star Odyssey: From Main Sequence to Supernova, 212, 230 

\reference{} Stevens, I.~R.\ 1995, \mnras, 
277, 163 

\reference{} Stevens, I.~R., 
Blondin, J.~M., \& Pollock, A.~M.~T.\ 1992, \apj, 386, 265 


\reference{} Tuthill, P.~G., 
Monnier, J.~D., \& Danchi, W.~C.\ 1999, \nat, 398, 487 


\reference{} Tuthill, P.~G., 
Monnier, J.~D., Lawrance, N., Danchi, W.~C., Owocki, S.~P., 
\& Gayley, K.~G.\ 2008, \apj, 675, 698 


\reference{} Underhill, A.~B., \& Hill, G.~M.\ 1994, \apj, 432, 770 

\reference{} Usov, V.~V.\ 1992, \apj, 389, 635 


\reference{} van der Hucht, K.~A.\ 
2001, New Astronomy Review, 45, 135 

\reference{} Van Loo, 2005, Ph.D. Thesis, Royal Observatory of Belgium


\reference{} White, R.~L.\ 1985, \apj, 289, 698 

\reference{} Williams, P.~M., van 
der Hucht, K.~A., Pollock, A.~M.~T., Florkowski, D.~R., van der Woerd, H., 
\& Wamsteker, W.~M.\ 1990, \mnras, 243, 662 

\reference{} Williams, P.~M., van der Hucht, K.~A., \& Spoelstra, T.~A.~T.\ 1994, \aap, 291, 805 



\reference{} Williams, P.~M., 
Dougherty, S.~M., Davis, R.~J., van der Hucht, K.~A., Bode, M.~F., 
\& Setia Gunawan, D.~Y.~A.\ 1997, \mnras, 289, 10 


\reference{} Wright, A.~E., \& Barlow, M.~J.\ 1975, \mnras, 170, 41 

\end{references}
\end{document}